\documentclass[aps,prl,twocolumn]{revtex4}
\usepackage{graphicx} 
\usepackage{dcolumn}
\usepackage{bm}
\usepackage{amsmath}
\usepackage{subfigure}
\usepackage{units}

\begin{document}

\title{Magnetic conveyor belt transport of ultracold atoms to a superconducting atomchip}

\author{S. Minniberger}
\affiliation{Vienna Center for Quantum Science and Technology, Atominstitut/TU-Wien, Stadionallee 2, 1020 Vienna, Austria}
\author{F. Diorico}
\affiliation{Vienna Center for Quantum Science and Technology, Atominstitut/TU-Wien, Stadionallee 2, 1020 Vienna, Austria}
\author{S. Haslinger}
\affiliation{Vienna Center for Quantum Science and Technology, Atominstitut/TU-Wien, Stadionallee 2, 1020 Vienna, Austria}
\author{C. Hufnagel}
\affiliation{Vienna Center for Quantum Science and Technology, Atominstitut/TU-Wien, Stadionallee 2, 1020 Vienna, Austria}
\author{Ch. Novotny}
\affiliation{Vienna Center for Quantum Science and Technology, Atominstitut/TU-Wien, Stadionallee 2, 1020 Vienna, Austria}
\author{N. Lippok}
\affiliation{Vienna Center for Quantum Science and Technology, Atominstitut/TU-Wien, Stadionallee 2, 1020 Vienna, Austria}
\author{J. Majer}
\affiliation{Vienna Center for Quantum Science and Technology, Atominstitut/TU-Wien, Stadionallee 2, 1020 Vienna, Austria}
\author{Ch. Koller}
\affiliation{Vienna Center for Quantum Science and Technology, Atominstitut/TU-Wien, Stadionallee 2, 1020 Vienna, Austria}
\author{S. Schneider}
\affiliation{Vienna Center for Quantum Science and Technology, Atominstitut/TU-Wien, Stadionallee 2, 1020 Vienna, Austria}
\author{J. Schmiedmayer}
\email{schmiedmayer@atomchip.org}
\homepage[URL: ]{http:\\www.atomchip.org}
\affiliation{Vienna Center for Quantum Science and Technology, Atominstitut/TU-Wien, Stadionallee 2, 1020 Vienna, Austria}

\date{\today}

\begin{abstract}
We report the realization of a robust magnetic transport scheme to bring $>3\times10^8$ ultracold $^{87}\unit{Rb}$ atoms into a cryostat. The sequence starts with standard laser cooling and trapping of $^{87}\unit{Rb}$ atoms, transporting first horizontally and then vertically through the radiation shields into a cryostat  by a series of normal- and superconducting magnetic coils. Loading the atoms in a superconducting microtrap paves  the way for studying the interaction of ultracold atoms with superconducting surfaces and quantum devices requiring cryogenic temperatures.
\end{abstract}

%\keywords{}
\maketitle

% First paragraph: Introduction
There has been growing interest in studying Hybrid Quantum Systems \cite{Schoelkopf2008k,Wallquist2009,Xiang2013,Rabl2006,Andre2008,SvdWCL:04,Petrosyan:08,Petrosyan:09,Ver09,Henschel2010,TaylorAtomInterface2012,Bernon2013, Amsuess2011,Wu2010, Schuster2010,Kubo2013,Imamoglu2009}.  Superconducting quantum devices are expected to be capable of fast quantum information processing but exhibit only short coherence times making them unsuitable for qubit storage \cite{Schoelkopf2008k}. Hybrid Quantum Systems promise that coupling to a different system with a long coherence time will allow for high fidelity storage of qubits. The hyperfine spin states in ultracold atoms are a promising candidate \cite{SvdWCL:04,Petrosyan:08,Petrosyan:09, Ver09,Henschel2010, TaylorAtomInterface2012,Bernon2013} which can be manipulated with great precision. For experiments with ultracold atoms and superconducting quantum devices, one must be able to efficiently trap ultracold atoms in a cryogenic environment.

% Second paragraph: Other transport experiments
Transport of ultracold atomic ensembles is a well established technique to separate the physical experiment from the initial preparation of an ultracold atomic ensemble. Since the first experiments with ultracold quantum gases aiming for Bose-Einstein condensation, several transport mechanisms based on moving magnetic traps \cite{Greiner2001} and moving optical lattices \cite{Denschlag2006} or optical tweezers \cite{Tweezers2002} were developed. Transfer of atoms into a 4K cryogenic environment has been demonstrated with either a single moving magnetic quadrupole trap \cite{mukai2007} or by operating a MOT in the cryogenic environment \cite{Roux08, Nirrengarten2006,Jessen2013}. We report a robust magnetic transport scheme for ultracold atoms from a room temperature MOT into a cryostat and successful loading into a superconducting atomchip microtrap.

% First figure: cut through the whole setup and the vertical current sequence
\begin{figure}[tb]
	\includegraphics[width=1\columnwidth]{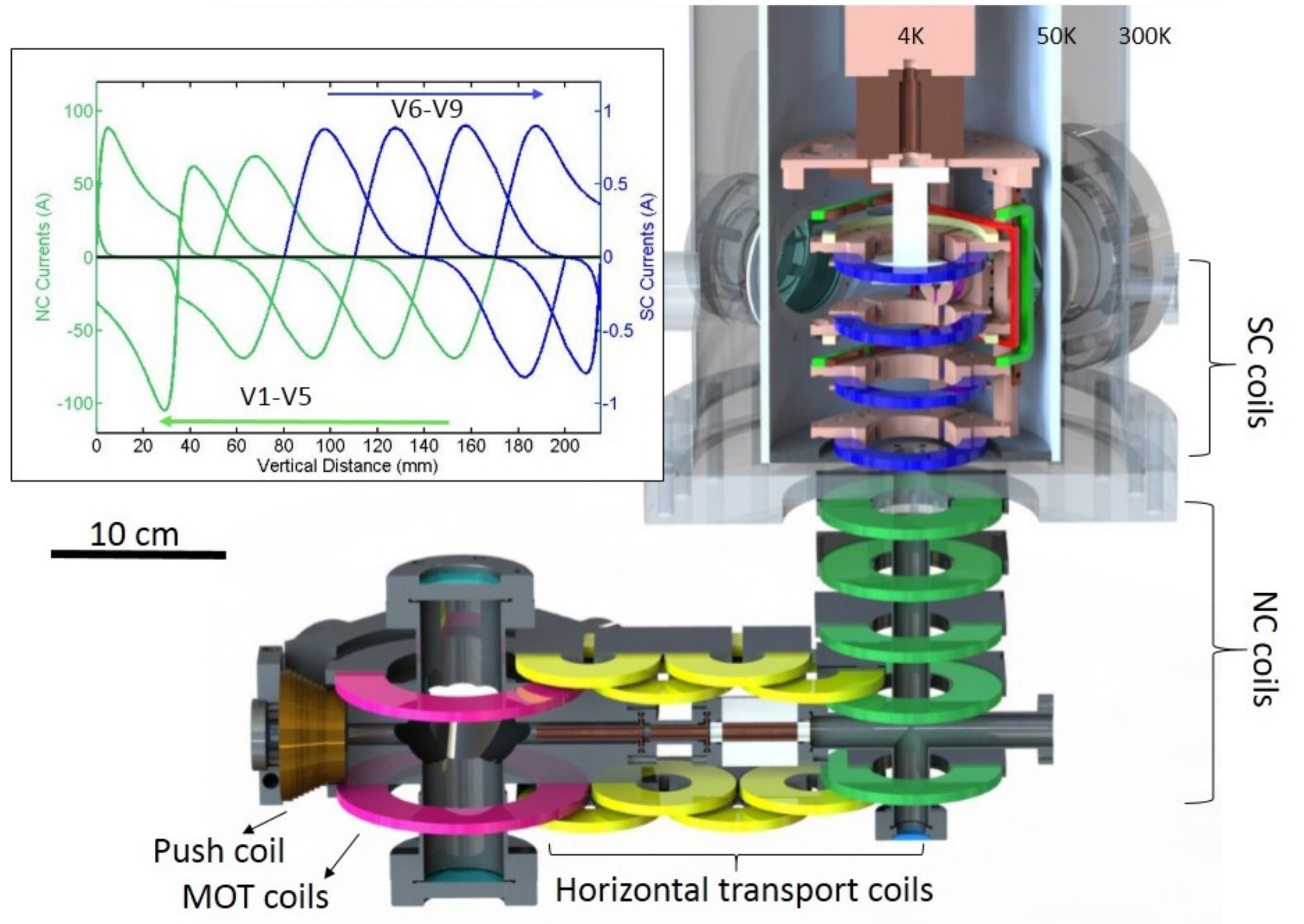}
	%\setbeamerfont{caption name}{size=\tiny}
	\caption{Cut through a CAD drawing of the setup with the MOT chamber on the bottom left and the copper cage holding the superconducting coils and the chip on the top right. The cryo chamber is made semi-transparent for clarity. The inset shows the currents used for the vertical transport. Currents for the normal conducting coils V1 to V5 refer to the left axis, while the superconducting coils V6 to V9 are driven with much lower currents (right axis).}
	\label{fig:cad}
\end{figure}

% Third paragraph: Overview, Vacuum setup
The experiment uses a combination of horizontal magnetic transport following \cite{Greiner2001} and vertical magnetic transport to enter the cryostat. Figure \ref{fig:cad} shows a drawing of the entire setup. A 90 degree angle in the path of the transport efficiently blocks stray light from the MOT. The non-overlapping arrangement of the vertical coils allows efficient transition into the cryogenic environment. The MOT chamber is separated by a valve, allowing independent maintenance of the science chamber.

% Fourth paragraph: Vacuum setup, cryostat, shield, optical access
The cryostat is an ultra-low vibration Gifford-McMahon closed cycle cryocooler \cite{Antohi2009,cryomodel}. It provides a cooling power of about $800\unit{mW}$ at the 4K stage. The cold finger rests in a CF200 vacuum chamber. An aluminum shield is connected to the 50K stage. It has four anti-reflection coated windows for optical access and shields the inner part from thermal radiation \cite{SF57}. To thermally isolate the 50K shield, its outside is wrapped with several layers of aluminized Mylar foil. Radiation shielding is particularly important for the superconducting wires, all of them are covered with reflective aluminum tape.

% Fith paragraph: Details about the MOT
The MOT chamber is a pancake shaped steel octagon which contains the $\unit{Rb}$-dispenser. It provides optical access for a standard six beam vapour cell MOT. Following a ten second MOT phase, the atoms are sub-Doppler cooled and then optically pumped into the strongest low field seeking state $|2,2\rangle$. The initial magnetic trap is loaded with typically $5\times10^8$ atoms at a temperature of about $300\unit{\mu K}$.

% Sixth paragraph: Details about the horizontal transport
To initiate the horizontal transport a so called "push coil" shifts the quadrupole trap created by the MOT coils towards the magnetic conveyor belt. Currents for the horizontal section are calculated by defining the zero of the quadrupole field along the transport axis, a constant vertical trap gradient of $130\unit{G/cm}$ and a constant aspect ratio of $1.62$.

% Seventh paragraph: Details about the vertical transport coils
The vertical transport section consists of nine coils in total, five normal- and four superconducting each with a vertical spacing of $30\unit{mm}$. The normal conducting coils, which are mounted on water cooled aluminum bodies, have 40 windings each and are operated up to $100\unit{A}$. For the superconducting coils, a commercial Niobium-Titanium (NbTi) wire with a thickness of $127\unit{\mu m}$ is used \cite{cryomodel}. Each coil has 3000 windings and is wound on a copper mounting, consisting of four isolated segments in order to prevent eddy currents. The superconducting coils can be operated up to $3\unit{A}$. 

% Seventh paragraph part II: Details about the currents
The last coil pair of the horizontal section also act as the first two vertical transport coils. To maintain a constant aspect ratio of the trap during transport, four coils are used at the same time. In contrast to the horizontal section, transport along the coil axis in vertical direction requires bipolar operation of the respective currents. For simulating the magnetic field of each coil, the analytic solution is used \cite{bfieldanal2001}. The currents $I_i(z)$ for coil $i$ and trap location $z$ along the z-axis are obtained using four conditions $|B(z)|=0$,  $|B_z|'=130\unit{G/cm}$, $|B_z|''=0$, and $\sum_{i=1}^{4} N_i \cdot I_i =0$ where $N_i$ is the number of windings per coil. The first three conditions imply a quadrupole trap with zero field at location $z$ and a linear gradient of $130\unit{G/cm}$. The last condition ensures smooth current over time. By specifying a position function $z(t)$ which contains the desired acceleration and maximum velocity, transport currents $I_i(t)$  are obtained from $I_i(z)$ \cite{lippokdiploma}.

% Seventh paragraph part III: transport speed
A maximum efficiency is achieved by using an acceleration of $0.4\unit{m/s^2}$ and a maximum velocity of $3\unit{m/s}$ for both horizontal and vertical transport. With these settings the whole magnetic transport sequence takes about two seconds \cite{Hasi2011}.

\begin{figure}[tb]
  	\subfigure[]{
  	\includegraphics[width=0.53\columnwidth]{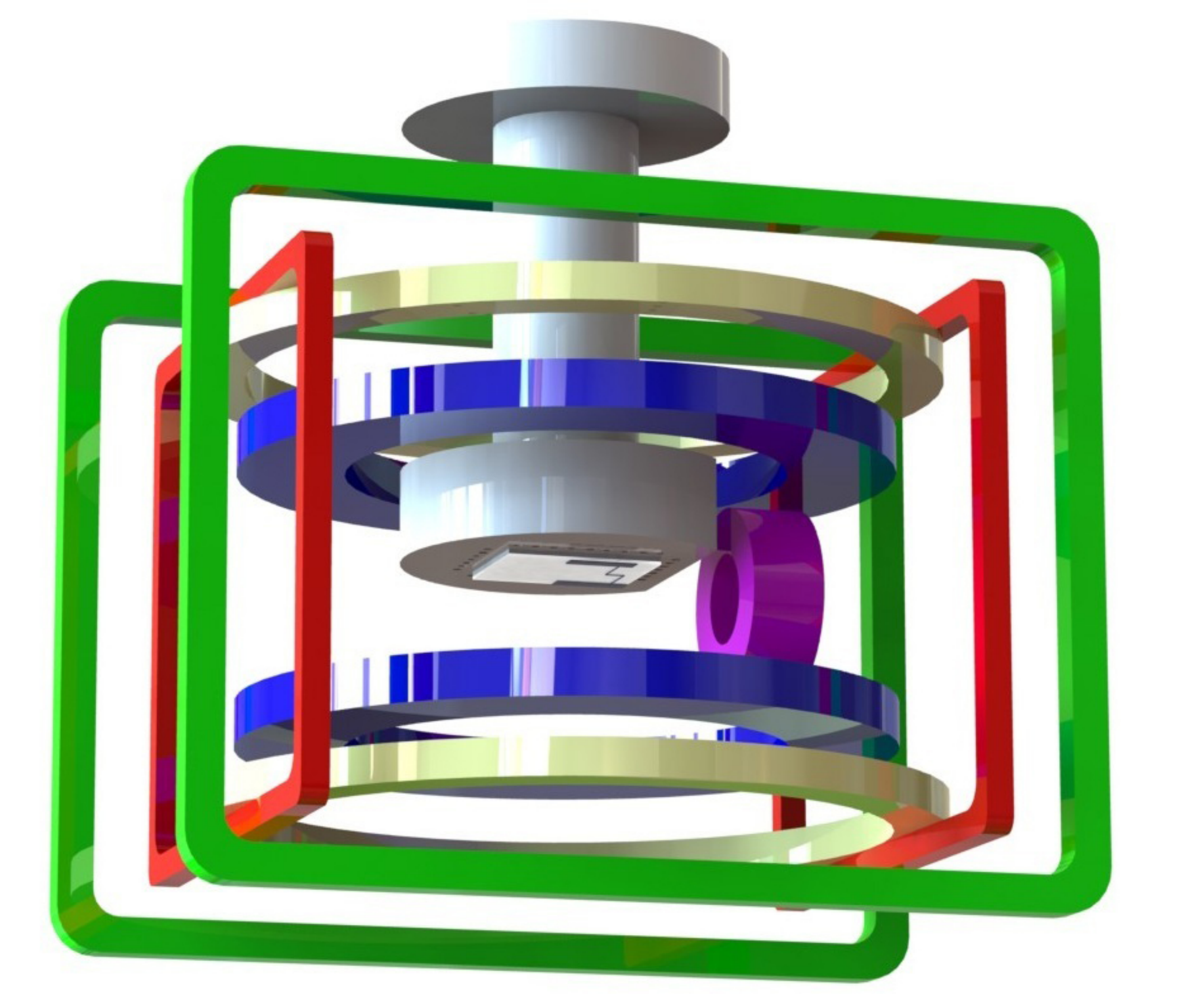}
  	\label{fig:cagerendering}
  	}
  	\subfigure[]{
  	\includegraphics[width=0.4\columnwidth]{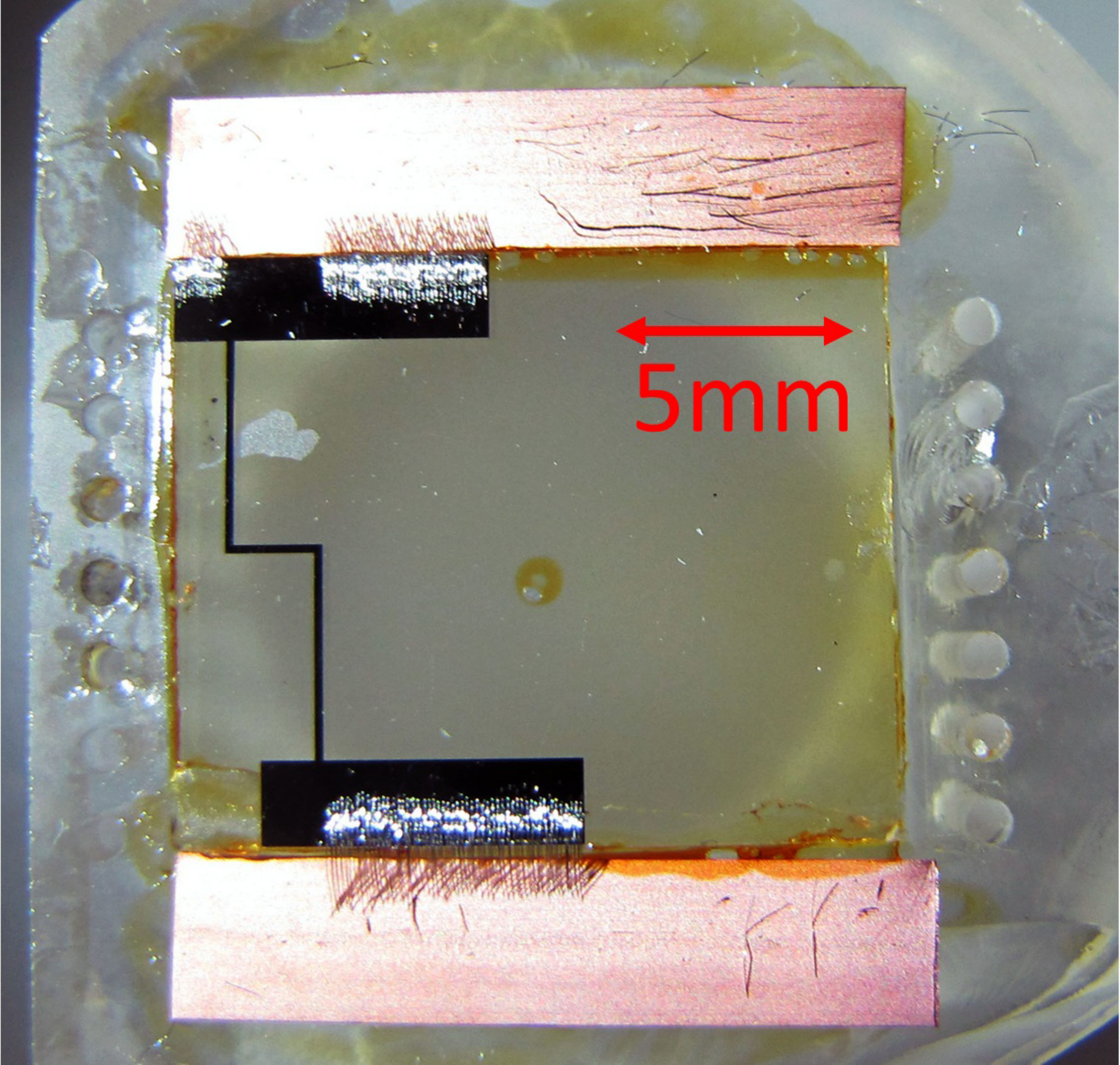}
  	\label{fig:chipphot}
  	} 	
  	\caption{(a) Coil configuration and the chip mounting in the cryostat, showing the last two transport coils (blue), the vertical bias coils (yellow), the bias coils for the chip trap (green), the bias coils for the third direction (red) and the Ioffe coil (small, pink). (b) Photo of the actual chip. Several aluminum bonds connect the Nb pads to the high-$T_c$  stripes.} 	
  	\label{fig:cagerendering_chipphot}
\end{figure}

% Eight paragraph: Details about the cage and the wiring
The main experimental stage is mounted on the 4K stage of the cold-finger. It consists of a copper cage system which holds the coils and the chip mount. The cryostat contains eleven coils in total, four transport coils, one Ioffe coil, and three coils pairs for homogeneous offset (bias) fields. The upper stage of the coil setup is shown in figure \ref{fig:cagerendering}. A total of 24 copper wires, capable of carrying up to 3A each, enter the cryostat. They are well thermally anchored to the 50K stage and 20K stage. To avoid heat dissipation, the wires are soldered to commercial high-$T_c$ coated superconductors which, subsequently, are connected to the NbTi wires at the 4K stage \cite{cryomodel}.

% Second figure: coilcage and chipphoto

% Ninth paragraph: Chip and mounting
The chip mounting is made of single crystal quartz to prevent eddy currents and still have a high thermal conductivity. The chip is made of a Sapphire substrate with a $500\unit{nm}$ thick sputtered niobium film ($T_c = 9.2\unit{K}$). A $100\unit{\mu m}$ wide Z-shaped wire with large contact pads is fabricated from this niobium layer with standard lithographic methods. To contact the niobium film, we use aluminum bonds between the contact pads and small pieces of the high-$T_c$ coated superconductors, which can then easily be soldered to the NbTi wires. A maximum current of 1.9A can be driven through this niobium wire structure, which corresponds to a current density of $3.8\times 10^6 \unit{A/cm^{2}}$. Figure \ref{fig:chipphot} shows the actual chip on the quartz mounting.

% Tenth paragraph: QUIC trap, precooling
At the end of the transport up to $3\times 10^8$ atoms at about $350\unit{\mu K}$ are held by the last two superconducting transport coils forming a quadrupole trap. This corresponds to a transport efficiency of about $60\%$, which is limited by the background pressure ($5 \times 10^{-9} \unit{mbar}$) in the room temperature part of the setup. In principle, the atomnumber in the cryostat can be increased by upgrading the MOT optics or improving the background pressure in the lower chamber. In the cryostat, the atomic clouds exhibit lifetimes of up to five minutes due to the low pressure in the cryogenic environment. After transport, the atoms are loaded into an intermediate trap using the vertical bias coils in anti-Helmholtz configuration. This allows the quadrupole trap to be connected in series with the Ioffe coil. This forms a Quadrupole-Ioffe configuration (QUIC) trap which minimizes heating due to current fluctuations \cite{Esslinger1998}. After switching, the intermediate trap is ramped down, the QUIC trap is ramped up to $I_{\unit{QUIC}}=1.2\unit{A}$. The atoms are now in a macroscopic trap with trapping frequencies of $(\omega_{\unit{long}},\omega_{\unit{radial}})/2\pi=25\unit{Hz},250\unit{Hz}$ and a trap bottom of $4\unit{G}$. A slow radio frequency ramp from $30\unit{MHz}$ to $5\unit{MHz}$ precools the atoms in the QUIC trap and increases the density of the cloud by evaporative cooling.

\begin{figure}[tb]
   	\subfigure[]{
   	\includegraphics[width=0.57\columnwidth]{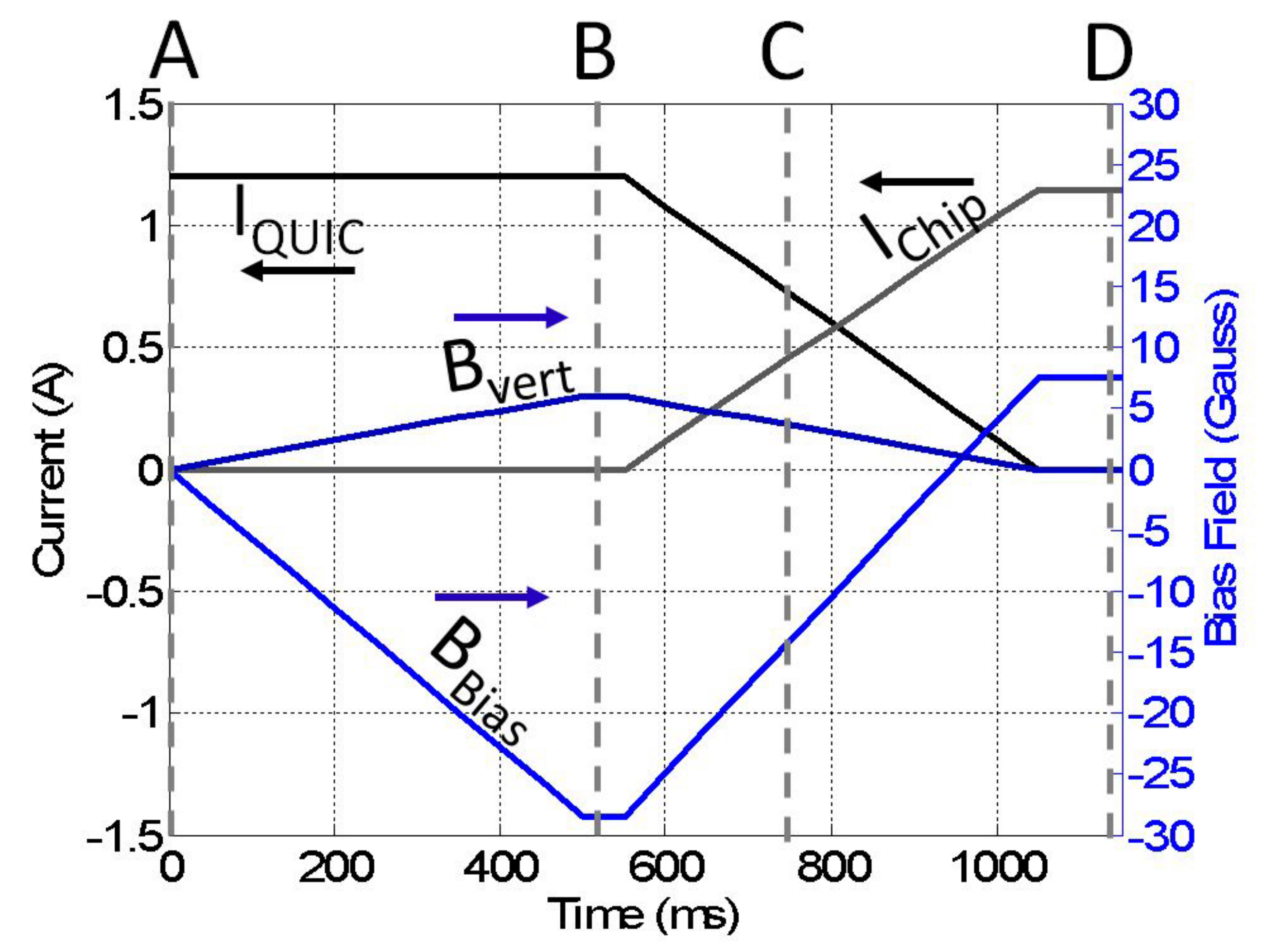}
   	\label{fig:sequence}
   	}
   	\subfigure[]{
   	\includegraphics[width=0.35\columnwidth]{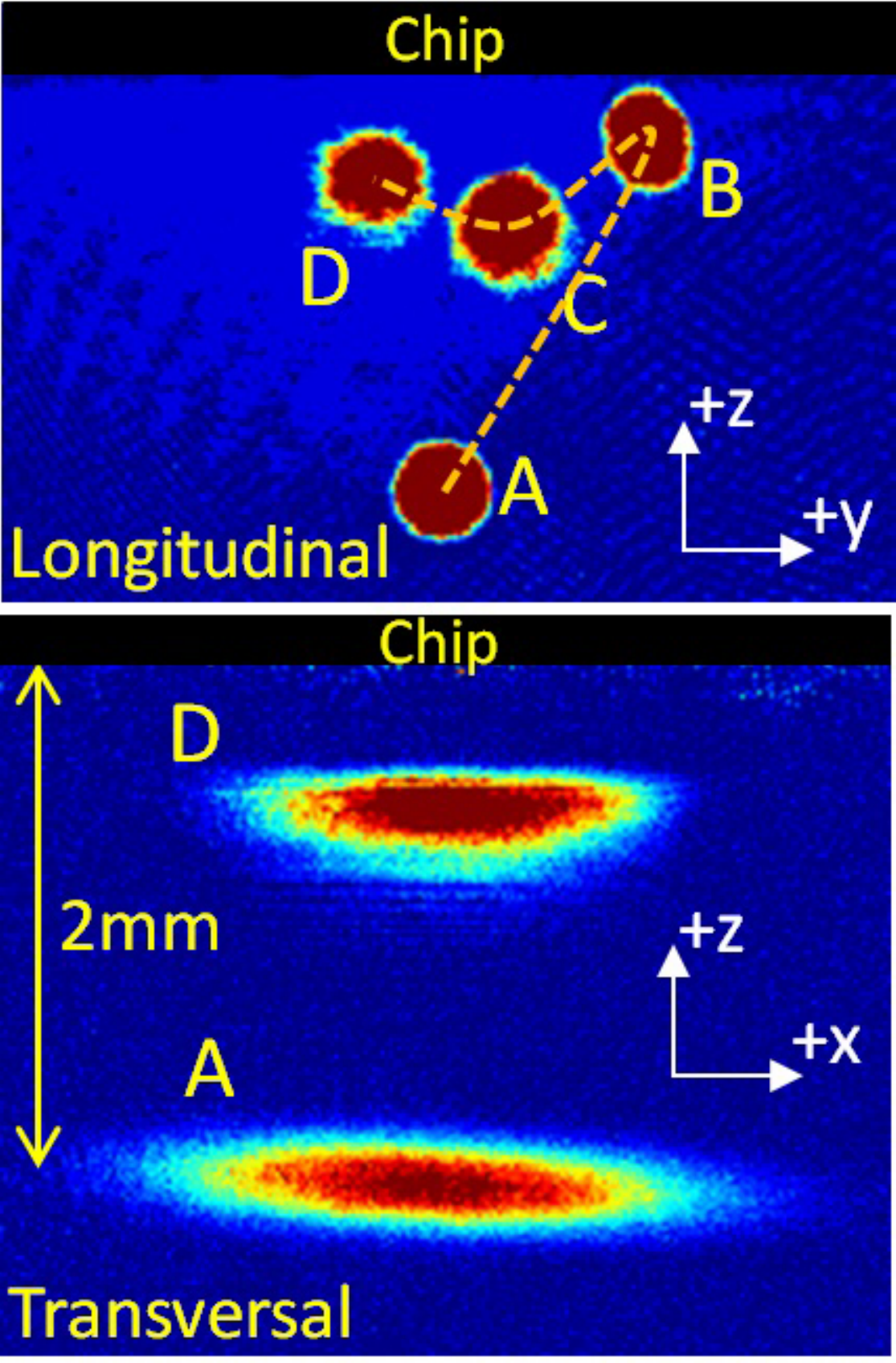}
   	\label{fig:atoms}
   	}
   	\caption{(a) Current sequence for loading, (b) Absorption images of the atoms taken along the longitudinal (top) and transversal (bottom) axes of the chip trap. Here, four images corresponding to different points throughout the loading sequence are stacked. A: initial QUIC trap, B: $-B_{Bias}$ and vertical offset at maximum, C: middle of the ramp and D: final chip trap. The dashed line shows the trajectory of the atoms through the entire loading sequence. } 	
   	\label{fig:sequence_atoms}
\end{figure}

% Eleventh paragraph: reloading and numbers for the chip trap
The transverse magnetic field of the Z-wire chip trap is rotated by $45^\circ$ relative to the corresponding field in the QUIC trap, which is fixed by the axis of the vertical transport. This makes direct loading difficult. This is overcome by implementing a "swing by maneuver". Figure \ref{fig:sequence} shows the whole loading sequence used to transfer the atoms from the QUIC trap into the superconducting Z-trap. The sequence starts by moving the QUIC trap closer to the chip with a vertical bias field, $B_{\unit{vert}}$, while at the same time moving it off-center with the chip bias field in opposite polarity, $-B_{\unit{bias}}$. The actual loading into the chip trap happens in the next step, when $I_{\unit{QUIC}}$ is ramped to zero, $I_{\unit{chip}}$ is ramped up and $-B_{\unit{bias}}$  is ramped to the actual bias field $+B_{\unit{bias}}$. This allows a smooth transition between the two rotated magnetic field configurations. The chip loading sequence was found to be optimal for a ramp time of $500\unit{ms}$. Figure \ref{fig:atoms} shows the atoms in the chip trap at different points during the transfer. The swing-by maneuver is best observed through the longitudinal direction, where the sidewards motion of the trap is visible. At the end of the sequence, the superconducting microtrap holds around $2\times10^6$ atoms at $20\unit{\mu K}$.  Using a current of $I_{\unit{chip}} = 1.12\unit{A}$ in the Z-wire and a bias field of $B_{\unit{bias}}=7.6\unit{G}$ results in a trap around $350 \unit{\mu m}$ from the chip surface. The measured trapping frequencies are $(\omega_{\unit{long}},\omega_{\unit{radial}})/2\pi=20\unit{Hz}, 370\unit{Hz}$  with a trap bottom of $2\unit{Gauss}$.
 
 % Third Figure: Reloading sequence
 
\begin{figure}[tb]
	 \includegraphics[width=0.95\columnwidth]{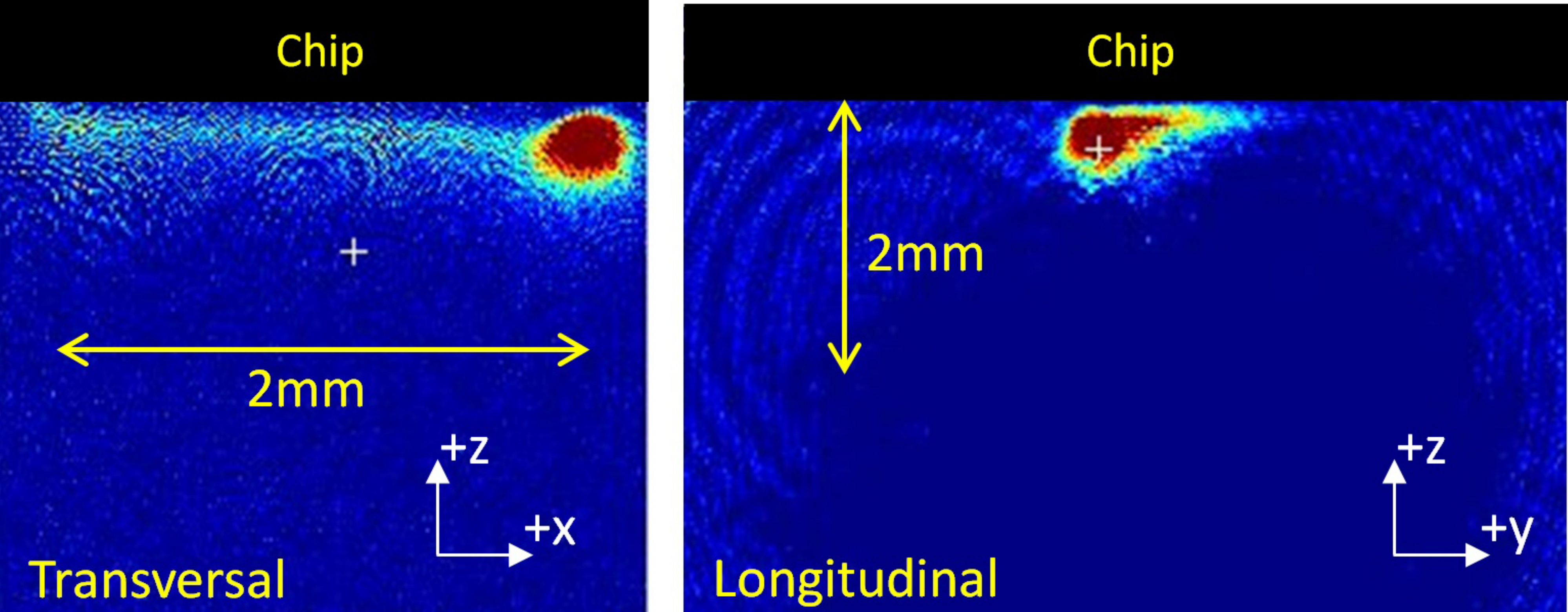}
	 \caption{Meissner trap formation after ramping down $I_{\unit{QUIC}}$ to zero while leaving only the vertical bias field on. Due to geometric deviations, the majority of the atoms are trapped on one of the Z-leads. The rest of the Z-structure is also weakly outlined by the Meissner trap.}
	 \label{fig:meissner}
 \end{figure}

% Twelvth paragraph: Meissner Trap
Bringing the trap closer than 2.5 times the width of the superconducting wire (here $100 \unit{\mu m}$), opens it up towards the chip surface due to the Meissner effect \cite{Dikovsky2008}. In fact, the Meissner effect can be used to trap the atoms by applying a vertical bias field \cite{Hufi2011}. This Meissner trap can easily be loaded directly from the QUIC trap. Figure \ref{fig:meissner} shows atoms in the Meissner trap along the superconducting surface of the Z-wire.

  % Fourth figure: Meissner trap
 
% 13th paragraph: Summary and outlook
We have successfully transported thermal atoms into a cryogenic environment and trapped them on a superconducting atomchip. The magnetic conveyor belt shows high robustness due to the absence of alignment sensitive parts and avoids the operation of a MOT close to superconducting surfaces. Furthermore, the turnaround time for modifications in the science chamber is less than a day due to cryo pumping. The transport scheme described here is fully compatible with a dilution refrigerator since they have similar cooling powers at the 4K stage.
This will enable experiments where ultracold atoms interact with superconducting quantum circuits. A superconducting microwave resonator can be integrated on the atomchip to study the coupling of ultracold atoms to microwave photons. Ultracold atoms near superconducting surfaces also open the possibility to use unique superconducting lattice traps \cite{vortexlattice}.

% 14th and last paragraph: Acknowledgements, author contributions
SM and FD contributed equally to the efforts leading to this paper. They would like to acknowledge the support of the COQUS doctoral program. This work was funded through the European Union Integrated Project SIQS and the Austrian Science Fund FWF (Wittgenstein Prize, SFB FOQUS). 

%\bibliography{ConveyorBeltVienna}{}
%\bibliographystyle{unsrt}
%\bibliographystyle{apsrev}
%\bibliographystyle{phaip}

\end{document}